\documentclass[aps,pre,twocolumn,notitlepage,superscriptaddress]{revtex4-1}

\usepackage{amsmath}
\usepackage{amssymb}
\usepackage{hyperref}
\usepackage{graphicx}
\usepackage[arrow, matrix, curve]{xy}
\usepackage{bbm}
\usepackage{bm}
\usepackage{yhmath}
\usepackage{color}
\usepackage{physics}
\usepackage{bbold}

\usepackage{color}

\renewcommand\vec[1]{\boldsymbol{\mathrm{#1}}}

\newcommand\diff{\mathrm{d}}

\usepackage[usenames,dvipsnames]{xcolor}
\hypersetup{colorlinks=true, linkcolor=BrickRed, urlcolor=blue!50!black, citecolor=blue!50!black}

\makeatletter
\newcommand\hide@visible[1]{%
  \bgroup\fboxsep=.3ex\colorbox{Gray}{begin hide}%
  #1\colorbox{Gray}{end hide}\egroup%
}
\newcommand\hide@hidden[1]{%
  \bgroup\fboxsep=.3ex\colorbox{Gray}{hidden text}%
}
\newcommand\hide@invisible[1]{}
\newcommand\makevisible{\let\hide\hide@visible}
\newcommand\makehidden{\let\hide\hide@hidden}
\newcommand\makeinvisible{\let\hide\hide@invisible}
\makeatother
\makehidden

\begin{document}

\title{Two-Dimensional Active Brownian Particles Crossing a Parabolic Barrier: \\ Finite Rectangular Domain with Absorbing Boundary Conditions}

\author{Michele Caraglio}
\email{Michele.Caraglio@uibk.ac.at}
\affiliation{Institut f\"ur Theoretische Physik, Universit\"at Innsbruck, Technikerstra{\ss}e 21A, A-6020, Innsbruck, Austria}

\date{\today}

\begin{abstract}
We solve the time-dependent Fokker-Planck equation for a two-dimensional active Brownian particle exploring a rectangular domain with absorbing boundary and in the presence of a parabolic barrier along one direction.
By taking those of a passive Brownian particle as basis states and dealing with the activity as a perturbation, we provide a matrix representation of the Fokker-Planck operator and express the propagator in terms of the perturbed eigenvalues and eigenfunctions.
Our solution also allows us to obtain the survival probability and the first-passage-time distribution.
The non-equilibrium character of the dynamics induces a strong dependence of the latter quantities on the particle's activity, while the rotational diffusivity influences them to a minor extent.
\end{abstract}

\maketitle

\section*{Introduction}

Active motion refers to the self-propulsion ability of certain natural and artificial microswimmers~\cite{Bechinger2016,Marchetti2013,Romanczuk2012,Elgeti2015}.
This peculiar characteristic assigns to active particles a relevant role in a wide variety of research fields~\cite{Needleman2017,Wang2012,Erkoc2019,Cates2012,Fily2012,Stenhammar2015,Volpe2014}.
However, even after more than two decades of intensive research, models of active particles have been analytically exactly solved only in few setups~\cite{Wagner2017,Hermann2018,Schnitzer1993,Tailleur2008,Tailleur2009,Malakar2018,Kurzthaler2016,Kurzthaler2017,Kurzthaler2018,Martens2012,Sevilla2015,Malakar2020}.

Recently, it has been shown that an exact solution of the Fokker-Planck equation of an active Brownian particle (ABP) can be obtained once the basis states of a reference standard Brownian particle are known and the activity is regarded as a perturbation of this reference system.
This method successfully applied to ABP in an isotropic harmonic potential~\cite{Caraglio2022}, in a circular region with an absorbing boundary~\cite{DiTrapani2023}, in front of an absorbing wall~\cite{Bauche2025}, and in the presence of a parabolic barrier with absorbing boundary conditions at both sides~\cite{Caraglio2025}.
In these papers, the solution of the Fokker-Planck equation amounts to a formally exact series expression for the probability propagator in terms of the perturbed eigenvalues and eigenfunctions.
In particular, in Ref.~\cite{Caraglio2025}, the focus was on the case of a domain infinitely extended in the direction perpendicular to that of the parabolic barrier and on deriving the transition-path-time statistics~\cite{Chung2009,Neupane2016,Zhang2007,Kim2015,Caraglio2018,Caraglio2020}.
Here, building on the results of Ref.~\cite{Caraglio2025}, we show that the above method can be further leveraged to investigate the behavior of an ABP navigating in a finite rectangular domain delimited by absorbing boundaries in each direction and characterized by a parabolic barrier along one of the coordinates.

The interest in the environment here considered is twofold:
On the one hand, a parabolic barrier represents a first approximation of any realistic potential barrier in the vicinity of its peak. 
Hence, being able to exactly describe the dynamics of a ABP over a parabolic barrier may provide a deeper understanding of its behavior in more complex potential landscapes. 
In particular, it can offer new insight into the escape dynamics of active particles, a topic that has been extensively investigated both experimentally~\cite{Militaru2021,Wen2023} and theoretically~\cite{Sharma2017,Caprini2019,Caprini2021,Zanovello2021}.
On the other hand, a finite domain provides a setup that better represents typical experimental situations, in which active particles are confined by the physical boundaries of the experimental apparatus or by ad-hoc placed hard walls, channels or confining potentials.
In particular, absorbing boundaries represent natural or experimental scenarios where particles are permanently removed upon reaching certain regions, such as chemical reactions at reactive surfaces or particles escaping confinement~\cite{Bressloff2023}, or in which meeting the boundary triggers certain experimental actions~\cite{Muinos-Landin2021}.
Furthermore, perfectly absorbing boundaries are a limiting case of sticky boundary conditions, which are a way to formulate the accumulation process at confining boundaries typically exhibited by both ABPs and Run-and-Tumble particles, even at the single particle level~\cite{Bressloff2023b}.
Finally, absorbing boundaries allow to develop a theoretical framework aiming at investigating first-passage properties~\cite{Redner2001}.

Indeed, by properly integrating the propagator, we can derive series expressions for the survival probability, the first-passage-time distribution, and the mean first-passage time~\cite{Redner2001,Metzler2013,Risken1989,Palyulin2019}.
Both these observables play a pivotal role in understanding several processes in nature, including topological disentanglement of knots on open polymers~\cite{Caraglio2019}, chemical processes such as fluorescence quenching~\cite{Chmeliov2013}, electrical impulses from a neuron~\cite{Gerstner1997}, and execution of buy/sell orders in financial markets~\cite{Sazuka2009,Baldovin2015}.
First-passage-time distribution is also exploited to characterize transport properties and escape dynamics of living micro-organisms or artificial nano- and micro-particles in different environments~\cite{Redner2001}.
However, while this observable has been widely investigated in the case of passive Brownian particles~\cite{Kim2015,Klein1952,Chatterjee2018,Durang2019,Besga2021}, less is known about its properties for self-propelled particles.
In particular, analytical expression of the mean and/or the distribution of first-passage time have been obtained only for one-dimensional run-and-tumble particles in several situations~\cite{Angelani2014,Demaerel2018,Malakar2018,Dhar2019}, for one-dimensional ABPs in a double well~\cite{Caprini2021}, in the long and short time regime for two-dimensional ABPs in a homogeneous environment~\cite{Basu2018}, and for ABPs in any dimension but only in the small noise limit and in confining potentials~\cite{Woillez2019}.
Furthermore, exploiting the knowledge of the propagator, we also derive expressions for the probability current and for the probability of absorption at a given boundary.
The latter observable is particularly useful for characterizing the dynamics of the system, as it provides direct information on how particles interact with the absorbing boundaries and on the relative likelihood of escaping through different sides of the domain.

Given the similarity of the problem, the structure and the phrasing of this manuscript deliberately follow that of Ref.~\cite{Caraglio2025}.
However, the problem considered in this manuscript can be considered as a generalization of that reported in Ref.~\cite{Caraglio2025} in the sense that the latter corresponds to the case in which the distance between the two absorbing boundaries perpendicular to the barrier is infinite.
Furthermore, in Ref.~\cite{Caraglio2025} we disregarded the spatial coordinate perpendicular to the direction of the barrier and we found a solution accounting only for the remaining coordinates, namely the coordinate along the direction of the barrier and the self-propulsion angle.
Here, the finiteness of the rectangular domain allows us to find a solution that takes into account both the Cartesian coordinates of the active particle, but this comes at the price that the eigenvalues and the eigenfunctions display three indices rather than only two as it was the case in the reduced problem solved in Ref.~\cite{Caraglio2025}.
We show that, as expected, depending on the initial conditions, the presence of the two absorbing boundaries perpendicular to the barrier leaves a clear fingerprint on the observables under investigation.

\section*{Model}

The stochastic overdamped motion of a two-dimensional ABP is completely characterized in terms of the propagator $\mathbb{P}(\vec{r}, \vartheta, t | \vec{r}_0 , \vartheta_0)$ which is the probability of finding the particle at position $\vec{r}=(x,y)$ and orientation $\vartheta$ at lag time $t$ given the initial position $\vec{r}_0$ and orientation $\vartheta_0$ at time $t=0$.

In the presence of a parabolic barrier along the $x$-direction, $U(x) = -k x^2/2$, with spring constant $k > 0$, the Fokker-Planck equation reads
\begin{align} \label{eq:eom_propagator}
    \partial_t\mathbb{P}  = \Omega \mathbb{P} := &  D \partial_x \left( e^{\mu k x^2 / 2D} \partial_x e^{-\mu k x^2 / 2D} \mathbb{P} \right)  \nonumber \\ 
    & + D \partial_y^2 \mathbb{P} + D_{\text{rot}} \partial_\vartheta^2 \mathbb{P} -  v \vec{u} \cdot \vec{\nabla} \mathbb{P} \; ,
\end{align}
where $D$ and $D_{\text{rot}}$ are the translational and rotational diffusion coefficient, respectively, whereas $\mu$ is the mobility of the particle.
The particle is endowed with a self-propulsion with fixed velocity $v$ along the orientation $\vec{u} = (\cos \vartheta, \sin \vartheta)$.
Given the initial condition
\begin{align} \label{eq:initial_condition}
    \mathbb{P}(\vec{r}, \vartheta, t=0 | \vec{r}_0 , \vartheta_0) =  \delta(\vec{r}-\vec{r}_0) \delta(\vartheta-\vartheta_0) \; ,
\end{align}
the previous equation readily provides the formal solution of the propagator
\begin{align} \label{eq:formal_solution} \mathbb{P}(\vec{r}, \vartheta, t | \vec{r}_0 , \vartheta_0) = e^{\Omega t} \delta(\vec{r}-\vec{r}_0) \delta(\vartheta-\vartheta_0) \; .
\end{align}

Here, we are interested in investigating the dynamics of an ABP conditioned to the presence of a rectangular domain with absorbing boundary at $x = \pm d_x$ and $y = 0, 2d_y$.
\begin{align} 
	\mathbb{P}(x=\pm d_x, y, \vartheta, t | \vec{r}_0 , \vartheta_0) = 0 \; , \label{eq:boundary_condition_x} \\
	\left\lbrace 
	\begin{array}{l}
		\mathbb{P}(x, y=0, \vartheta, t | \vec{r}_0 , \vartheta_0) = 0 \; , \\
		\mathbb{P}(x, y=2d_y, \vartheta, t | \vec{r}_0 , \vartheta_0) = 0 \; , \\
	\end{array} \right.  \label{eq:boundary_condition_y}
\end{align}
and the further requirement that the initial position $\vec{r}_0$ is chosen inside the domain.

We exploit the distance $d := d_x$ to fix the length unit of the problem.
Taking the passive Brownian particle ($v=0$) as a reference, it is convenient to define the time unit $\tau$ as the typical time required by such a particle to cover the distance between the two boundaries along the $x$ direction, $\tau := d^2/D$.
Finally, the ratio $D/\mu = k_B T$ defines the energy unit and introduces an effective temperature that for a passive particle corresponds to the temperature of
the bath.
Besides the stiffness of the parabolic barrier $\beta k d^2$, our model has three other independent dimensionless parameters:
The P\'eclet number, $\text{Pe} := v \tau /d$, assessing the importance of the self-propulsion with respect to the diffusive motion, the \textit{``rotationality''}, $\gamma :=  D_{\text{rot}} \tau$,  measuring the magnitude of the rotational diffusion, and a form factor $\alpha := d_y/d$ characterizing the shape of the rectangular domain.

In order to find an expression for the propagator that is a solution of Eq.~\eqref{eq:eom_propagator}, first we make a time-separation ansatz for the propagator, $\mathbb{P}=\mathcal{E}(t) p(x) \psi(\vec{r},\vartheta)$.
Here we defined the Boltzmann weight $p(x) \propto e^{-\beta U(x)}$ with $\beta = 1/k_B T$ the inverse temperature, which adopting a convenient normalization reads
\begin{align} \label{eq:peq}
p(x) = \dfrac{\exp(\beta k x^2/2)}{2 \pi \alpha d} \; .
\end{align}
It is convenient to define a new operator $\mathcal{L}$ by splitting off the Boltzmann weight
\begin{align} \label{eq:splitting}
\Omega p(x) \psi(\vec{r},\vartheta) =: p(x) \mathcal{L} \psi(\vec{r},\vartheta)\; .
\end{align}

Inserting into~\eqref{eq:eom_propagator} and using the defined units yields
\begin{align}\label{eq:SE3}
    \frac{1}{\mathcal{E}}\partial_t \mathcal{E} = \dfrac{1}{\psi} \mathcal{L} \psi  = -\lambda \;,
\end{align}
where the last equal sign holds since the first and second term of the equation are functions of independent variables, and therefore can only be equal to each other if their value is independent of all variables.
We can now explicitly write the solution for the time-dependent component of the propagator, $\mathcal{E}(t)=\exp(-\lambda t)$, and proceed to solve the equation for the spatial and angular components only, which now reads
\begin{equation} \label{eq:eigenvalueproblem}
    \mathcal{L} \psi + \lambda \psi = 0 \; .
\end{equation}
The similarity to a quantum mechanical problem suggests to tackle the problem by dealing with the activity as a perturbation of the passive system.
We thus split the operator $\mathcal{L}$ into a passive contribution $\mathcal{L}_0$ and an active driving $\mathcal{L}_1$ according to
\begin{equation}
    \mathcal{L} = \mathcal{L}_0 + \text{Pe} \, \mathcal{L}_1 \; ,
\end{equation}
with
\begin{gather}
    \mathcal{L}_0 \psi= 
    \dfrac{1}{\tau} \left[ d^2 \partial^2_x + \beta k d^2 x \partial_x + d^2 \partial_y^2  + \gamma \partial_\vartheta^2 \right] \psi \;, \label{eq:L0} \\
    \mathcal{L}_1\psi=-\dfrac{d}{\tau} \left[\beta k x \cos \vartheta + \cos \vartheta \, \partial_x  +  \sin \vartheta \, \partial_y  \right] \psi \; . \label{eq:L1}
\end{gather}

\section*{Solution of the passive reference system}

To solve the unperturbed eigenvalue problem 
\begin{equation} \label{eq:SE4}
    \mathcal{L}_0 \psi = - \lambda \psi \; ,
\end{equation}
subjected to the initial~\eqref{eq:initial_condition} and the boundary~(\ref{eq:boundary_condition_x}-\ref{eq:boundary_condition_y}) conditions, first we decompose the three degrees of freedom into different $(n,m,s)$ modes with the ansatz
\begin{equation} \label{eq:eigenfunctions_ansatz}
    \psi_{n,m,s}(\vec{r},\vartheta)= e^{i s \vartheta} \sin \left(  m \pi \dfrac{ y}{2 \alpha d} \right)   \mathcal{X}_{n}(x) \; .
\end{equation}
In writing the previous ansatz we exploited the known solution of the standard one-dimensional diffusion problem $\partial^2_y f_m(y) = -a_m^2 f_m(y)$ with absorbing boundary conditions on the domain $[0,L]$, which is $ f_m(y) \propto \sin(m \pi y / L)$, with $m=1,2,\ldots$.

Inserting Eq.~\eqref{eq:eigenfunctions_ansatz} into~(\ref{eq:SE4}) yields the equation for the $x$ component
\begin{align} \label{eq:Xcomp0}
	\left( \partial^2_x + \beta k x \partial_x    \right) \mathcal{X}_n +   \beta k  \sigma_n   \mathcal{X}_n = 0 \; ,
\end{align}
where 
\begin{equation} \label{eq:EVa}
    \lambda_{n,m,s}= \dfrac{1}{\tau} \left[ \beta k d^2 \sigma_n +  \left( \dfrac{m \pi}{2 \alpha} \right)^2   + \gamma s^2 \right] \; .
\end{equation}
As seen in Ref.~\cite{Caraglio2025}, Equation~\eqref{eq:Xcomp0} has even and odd solutions given by
\begin{align}
	\mathcal{X}_n(x) = e^{-\beta k x^2 / 4} Y_n(x) \; ,
\end{align}
with
\begin{align} \label{EQ:Solutions_2}
Y_n (x) \!=\! \!
\left\lbrace \!\!\!
\begin{array}{l} 
 e^{-\beta k x^2 / 4} {_1\!}F_1 \!\! \left(  \dfrac{1}{2} \!-\! \dfrac{\sigma_n}{2} \!; 
\dfrac{1}{2} \!; \dfrac{\beta k x^2}{2} \right) 
\quad n\!=\!0,2,4,\ldots \\
\; \\
\sqrt{\beta k}\, x e^{-\beta k x^2 / 4} \, {_1\!}F_1 \!\! \left( \!  1 \!-\! \dfrac{\sigma_n}{2}  \!; 
\dfrac{3}{2} \!; \dfrac{\beta k x^2}{2} \! \right)
\: n\!=\!1,3,\ldots
\end{array}
\right. 
\end{align}
where $\,_1F_1(a;b;z)$ is the Kummer confluent hypergeometric function and the boundary conditions $Y_n (\pm d) = 0$ fix the allowed values of $\sigma_n$.

The explicit expression of the eigenfunctions reads
\begin{align} \label{eq:eigenfunctions}
    \psi_{n,m,s}(\vec{r}, & \vartheta)  =  e^{i s \vartheta} \sin \left(  m \pi \dfrac{ y}{2 \alpha d} \right)  \dfrac{e^{-\beta k x^2/4} Y_n(x)}{\mathcal{N}_n} \; ,
\end{align}
where the normalization constant has been chosen such that
\begin{align} \label{eq:normalization}
    \braket{\psi_{n',m',s'}}{\psi_{n,m,s}} = \delta_{n,n'} \, \delta_{m,m'} \, \delta_{s,s'} \; .
\end{align}
Here we introduced the Kubo scalar product
\begin{align} \label{eq:KuboScalarProduct}
    \braket{\phi}{\psi} := \int_{-d}^{d} \! \diff x \int_{0}^{2 \alpha d} \! \diff y \int_0^{2\pi} \! \diff \vartheta \, p(x) \phi(\vec{r},\vartheta)^* \psi(\vec{r},\vartheta) \; ,
\end{align}
and resorted on the fact that the functions $Y_n(x)$ are orthogonal and normalizable with
\begin{align} \label{EQ:Y_normalization}
\int_{-d}^d \!\!  \diff x \, Y_n^2 (x) = \mathcal{N}_n^2 \; .
\end{align}
The isomorphism between $\ket{\psi}$ and $\psi(\vec{r},\vartheta)$ is made explicit by introducing generalized position and orientation states $|\vec{r} \vartheta\rangle$ such that $\psi(\vec{r},\vartheta) = \langle \vec{r}\vartheta |\psi \rangle$.
Using the orthogonality condition~\eqref{eq:normalization} it is easy to see that the operators $\ket{\psi_{n,m,s}} \bra{\psi_{n,m,s}}$ are a set of orthogonal projectors and thus we can write the following identity relation
\begin{align} \label{eq:completeness_Hilbert_space}
    \sum_{n,m,s} \ket{\psi_{n,m,s}} \bra{\psi_{n,m,s}} = \mathbb{1} \; ,
\end{align}
where we introduced a compact notation for the summation
\begin{align}
    \sum_{n,m,s}  :=  \sum_{n=0}^\infty \sum_{m=1}^\infty \sum_{s=-\infty}^\infty  \; .
\end{align}

The eigenfunctions of the passive reference system fulfill the completeness relation
\begin{align} \label{eq:completeness}
   p(x) \!\! \sum_{n,m,s} \!\! \psi_{n,m,s}(\vec{r},\vartheta) \psi_{n,m,s}(\vec{r}_0,\vartheta_0)^* \!=\!  \delta(\vec{r} \!-\! \vec{r}_0) \delta(\vartheta \!-\! \vartheta_0) \; .
\end{align}

The previous completeness relation~\eqref{eq:completeness} allows us to find a solution for the propagator in the passive reference system starting from its formal expression~\eqref{eq:formal_solution}
\begin{align}\label{eq:solution_propagator}
    \mathbb{P}^{(0)} & (\vec{r}, \vartheta, t | \vec{r}_0 , \vartheta_0)  =   e^{\Omega t} \delta(\vec{r}-\vec{r}_0) \delta(\vartheta-\vartheta_0)   \nonumber \\
    & = p(x) \sum_{n,m,s} \!\! \left\lbrace e^{\mathcal{L}_0 t} \psi_{n,m,s}(\vec{r},\vartheta)  \right\rbrace \psi_{n,m,s}(\vec{r}_0,\vartheta_0)^* \nonumber \\ 
    & = p(x) \sum_{n,m,s} \bra{\vec{r}\vartheta} e^{\mathcal{L}_0 t} \ket{ \psi_{n,m,s}} \braket{\psi_{n,m,s}}{\vec{r}_0 \vartheta_0} \nonumber \\
    & = p(x) \sum_{n,m,s} e^{-\lambda_{n,m,s}t} \, \psi_{n,m,s}(\vec{r}_0 ,\vartheta_0)^* \, \psi_{n,m,s}(\vec{r} ,\vartheta)
\; .
\end{align}
Note that, from the second line of the previous equation and using the identity relation~\eqref{eq:completeness_Hilbert_space}, one can also write
\begin{align}\label{eq:solution_propagator_2}
    \mathbb{P}^{(0)} &(\vec{r}, \vartheta, t | \vec{r}_0 , \vartheta_0)  = p(x) \bra{\vec{r}\vartheta} e^{\mathcal{L}_0 t} \ket{\vec{r}_0 \vartheta_0}
\; ,
\end{align}
meaning that the propagator is the projection of the generalized position and orientation state $\ket{\vec{r}\vartheta}$ over the time evolution of the initial state $\ket{\vec{r}_0\vartheta_0}$, multiplied by the Boltzmann weight $p(x)$.

\section*{Solution for ABP particles}

One readily shows that the passive operator $\mathcal{L}_0$ is Hermitian, $\braket{\phi}{\mathcal{L}_0\psi} = \braket{\mathcal{L}_0 \phi}{\psi}$, with respect to the Kubo scalar product~\eqref{eq:KuboScalarProduct} and consequently its eigenvalues $\lambda_{n,m,s}$ are real and left and right eigenfunctions  coincide, $\ket{\psi_{n,m,s}^{\text{L}}} = \ket{\psi_{n,m,s}^{\text{R}}} =\ket{\psi_{n,m,s}}$.
However, the full operator $\mathcal{L}$ does not reflect this property.
Correspondingly, in the following one has to be careful that the eigenvalues of the full operator, $\lambda_{n,m,s}^{\text{Pe}}$ are in general complex and the left eigenfunctions, $\ket{\psi_{n,m,s}^{\text{Pe},\text{L}}}$, are distinct from the right ones $\ket{\psi_{n,m,s}^{\text{Pe},\text{R}}}$.
If properly normalized, the perturbed left and right eigenfunctions constitute a bi-orthonormal basis with identity relation
\begin{align} \label{eq:completeness_Hilbert_space_perturbed}
    \sum_{n,m,s} \ket{\psi_{n,m,s}^{\text{Pe},\text{R}}} \bra{\psi_{n,m,s}^{\text{Pe},\text{L}}} = \mathbb{1} \; ,
\end{align}
which directly yields the propagator of the full problem
\begin{align}\label{eq:solution_propagator_full_problem}
    \mathbb{P} &(\vec{r}, \vartheta, t | \vec{r}_0 , \vartheta_0)  =  \bra{\vec{r}\vartheta} e^{\Omega t} \ket{\vec{r}_0 \vartheta_0} \nonumber \\
    & = p(x) \sum_{n,m,s} \bra{\vec{r}\vartheta} e^{\mathcal{L} t} \ket{ \psi_{n,m,s}^{\text{Pe},\text{R}}} \braket{\psi_{n,m,s}^{\text{Pe},\text{L}}}{\vec{r}_0 \vartheta_0} \nonumber \\
    & = p(x) \sum_{n,m,s} e^{-\lambda_{n,m,s}^{\text{Pe}}t} \, \psi_{n,m,s}^{\text{Pe},\text{L}}(\vec{r}_0 ,\vartheta_0)^* \, \psi_{n,m,s}^{\text{Pe},\text{R}}(\vec{r} ,\vartheta)
\; .
\end{align}

To explicitly compute the full propagator~\eqref{eq:solution_propagator_full_problem}, it is then necessary to calculate the perturbed eigenvalues and left and right eigenfunction.
To this scope, one has first to explicitly evaluate the action of the perturbation $\mathcal{L}_1$ on the eigenstates of $\mathcal{L}_0$.
Starting from Eqs.~\eqref{eq:L1} and~\eqref{eq:eigenfunctions} is it possible to show that
\begin{align}\label{eq:L1action}
\mathcal{L}_1  & \ket{\psi_{n,m,s}}  \!=\! -\dfrac{d}{\tau} \Bigg\{  \dfrac{1}{2} \sum_{n'=0}^{\infty} b_{n,n'} \big( \ket{\psi_{n',m,s+1}} + \ket{\psi_{n',m,s-1}}\big) \nonumber \\
 & - \dfrac{m \pi i}{4 \alpha d} \sum_{m'=0}^{\infty} a_{m,m'} \big( \ket{\psi_{n,m',s+1}} - \ket{\psi_{n,m',s-1}}\big)  \Bigg\} \; ,
\end{align}
with weights $a_{m,m'}$ and $b_{n,n'}$ having a rather lengthy formula which can be found in \hyperref[sec_appA]{appendix A}.

Now, given the finite-dimensional subspace of passive system's eigenfunctions such that $0 \leq n \leq n_{\text{max}}$, $0 < m \leq m_{\text{max}}$, and $|s| \leq s_{\text{max}}$, the action of the full operator $\mathcal{L}=\mathcal{L}_0 + \text{Pe} \,\mathcal{L}_1$ is completely characterized by a square matrix $\mathcal{G}$ of dimension $(1+n_{\text{max}}) m_{\text{max}} (2s_{\text{max}} + 1)$ with elements defined by
\begin{align}
& [\mathcal{G}]_{ (n' m_{\rm max} \!+\! m'\!-\!1) (2s_{\rm max} \!+\! 1) \!+\! s' \!+\!  s_{\rm max} , (n m_{\rm max} \!+\! m \!-\! 1) (2s_{\rm max} \!+\! 1) \!+\! s \!+\! s_{\rm max} } \nonumber \\
& \qquad \qquad = \bra{\psi_{n',m',s'}} \mathcal{L}_0 + \text{Pe} \, \mathcal{L}_1 \ket{\psi_{n,m,s}} \, ,
\end{align}
which has to be diagonalized numerically to obtain its eigenvalues $\lambda_{n,m,s}^{\text{Pe}}$ and left and right eigenvectors, $\bra{\psi_{n,m,s}^{\text{Pe},\text{L}}}$ and $\ket{\psi_{n,m,s}^{\text{Pe},\text{R}}}$ for any arbitrary P\'eclet number.
The perturbed eigenvectors are then a linear combination of the passive system's eigenstates
\begin{align}
\ket{\psi_{n,m,s}^{\text{Pe},\text{R}}} & = \sum_{n',m',s'} g_{n,m,s}^{\text{R};\, n',m',s'} \ket{\psi_{n',m',s'}} \; , 
\label{eq:decomposition_eigenstates1} \\
\bra{\psi_{n,m,s}^{\text{Pe},\text{L}}} & = \sum_{n',m',s'} g_{n,m,s}^{\text{L};\, n',m',s'} \bra{\psi_{n',m',s'}} \; . \label{eq:decomposition_eigenstates2}
\end{align}

\begin{figure}[t!]
\centering
\includegraphics[scale=1]{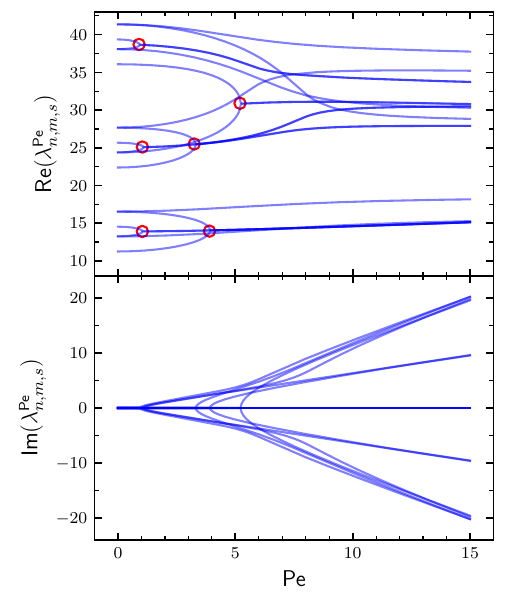} 
\caption{Numerical eigenvalues $\lambda_{n,m,s}$ of the Fokker-Planck operator $\mathcal{L} = \mathcal{L}_0+\text{Pe} \, \mathcal{L}_1$ as a function of the P{\'e}clet number $\text{Pe}$, for $\beta k d^2 = 10$, $\gamma=2$, $\alpha=1.5$, $n_{\text{max}}=2$, $m_{\text{max}}=2$, and $s_{\text{max}}=1$. Transparency of lines and exceptional points highlighted with red circles better show when real components merge and imaginary ones bifurcate.  \label{fig:spectrum}}
\end{figure}

The computational effort required to diagonalize such matrices increases rapidly with their dimension. 
However, the decaying exponentials in time in the expression of the propagator, Eq.~\eqref{eq:solution_propagator_full_problem}, ensures convergence.
In the unperturbed case, $\text{Pe}=0$, the eigenvalues~\eqref{eq:EVa} are real and an increasing function of $n$, $m$ and $|s|$.
For $\text{Pe} \neq 0$ the matrix corresponding to the full operator has certain out-of-diagonal elements having a complex value whose imaginary part becomes more and more important with decreasing $\alpha$.
This is expected since for small $\alpha$ the solution should be dominated by absorption at the upper and lower boundaries.
Nevertheless, as tested for values of $\alpha$ as small as $10^{-3}$, this does not pose a problem when calculating the eigenvalues as their complex component becomes significant only when, with increasing P{\'e}clet number, the real components of two eigenvalues merge and they bifurcate to a pair of complex conjugates for even larger activity, see Fig.~\ref{fig:spectrum}.
These branching points, called exceptional points~\cite{Heiss2012}, often originate in parameter-dependent eigenvalue problems and occur in a great variety of physical problems including mechanics, electromagnetism, atomic and molecular physics, quantum phase transitions, and quantum chaos. 
They have also been observed in other problems concerning active particles~\cite{Kurzthaler2016,Kurzthaler2017,DiTrapani2023}.
The exceptional points are highlighted with red circles in the upper panel of Fig.~\ref{fig:spectrum}.

\begin{figure*}[t!]
\centering
\includegraphics[scale=1]{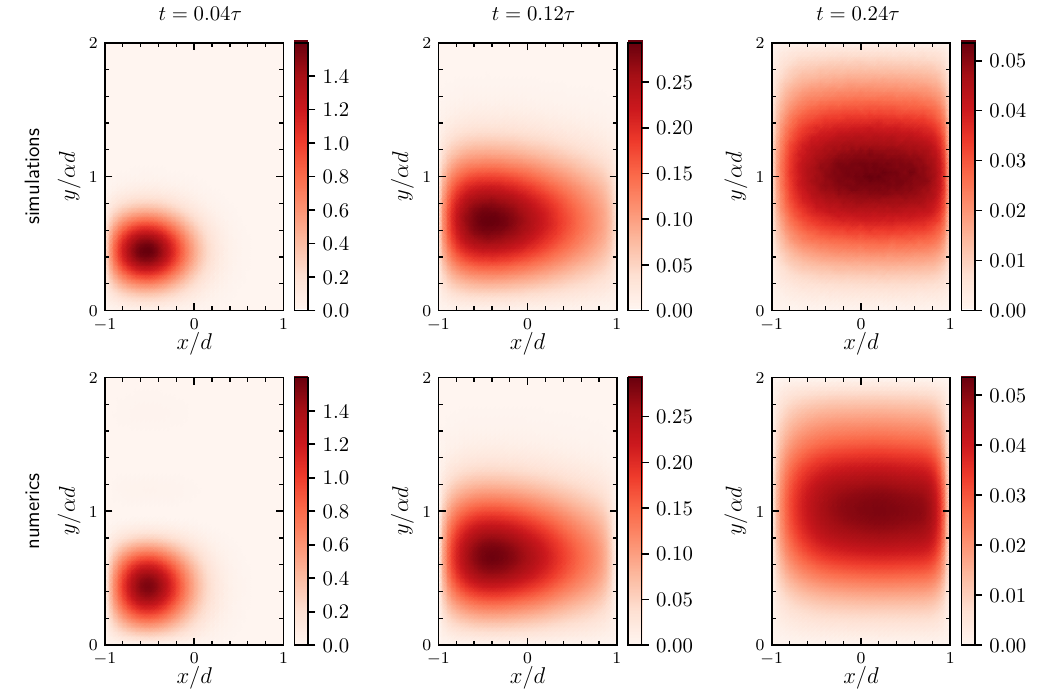}
\caption{Spatial probability distribution at different times $t$ starting with initial condition $x_0 = -d/2$, $y_0 = 0.5d$, and $\vartheta_0 = \pi/4$. Comparison between simulations, numerics for $\beta k d^2 =10$, $\alpha=1.5$, $\text{Pe} = 6$ and $\gamma=0.2$.
For the simulations, statistics has been collected from $10^8$ independent particles.
For the numerics, $n_{\text{max}}=6$, $m_{\text{max}}=7$, and $s_{\text{max}}=5$.
In both cases, the distributions were obtained by binning each spatial direction with a resolution of $0.05d$, yielding $41$ and $61$ bins along the $x$ and $y$ axes, respectively.\label{fig:spatial_probability}}
\end{figure*}

To corroborate our findings, we benchmark the time evolution of the spatial probability distribution starting from some given initial condition as obtained from numerics against that obtained by direct stochastic simulations, see Fig.~\ref{fig:spatial_probability}.
Details on the numerical simulations are reported in \hyperref[sec_appB]{appendix B}.
Note once more that, even if our analytical solution is formally exact, its numerical implementation requires truncating the series expansion appearing in the expression of the propagator, Eq.~\eqref{eq:solution_propagator_full_problem}. 
Thus, comparison with numerical simulations also serves as a sanity check to ensure that we have chosen sufficiently large values of $n_{\rm max}$, $m_{\rm max}$, and $s_{\rm max}$.

\section*{First order expansion of the propagator}

To fully exploit the perturbation theory formalism, we can perform a power-series expansion in the P\'eclet number of the perturbed eigenvalues and eigenfunctions.
\begin{align}
\lambda_{n,m,s}^{\text{Pe}} & =  \sum_{q=0}^{\infty}  \text{Pe}^q \, \lambda_{n,m,s}^{(q)}  \; ,  \label{eq:lambdaexpansion} \\
\ket{\psi_{n,m,s}^{\text{Pe},\text{R}}} & = \sum_{q=0}^{\infty}  \text{Pe}^q \, \ket{\psi_{n,m,s}^{\text{R},(q)}}  \; , \label{eq:psiexpansion} \\
\bra{\psi_{n,m,s}^{\text{Pe},\text{L}}} & = \sum_{q=0}^{\infty}  \text{Pe}^q \, \bra{\psi_{n,m,s}^{\text{L},(q)}}  \; ,
\end{align}
with $\lambda_{n,m,s}^{(0)} = \lambda_{n,m,s}$ and $\ket{\psi_{n,m,s}^{\text{R},(0)}} = \ket{\psi_{n,m,s}}$, and $\bra{\psi_{n,m,s}^{\text{L},(0)}} = \bra{\psi_{n,m,s}}$.
Focusing on the right perturbed eigenvector, inserting Eqs.~\eqref{eq:lambdaexpansion} and~\eqref{eq:psiexpansion} in the eigenvalue problem, Eq.~\eqref{eq:eigenvalueproblem}, yields
\begin{align}
\left( \mathcal{L}_0 + \text{Pe} \, \mathcal{L}_1 \right) & \sum_{q=0}^{\infty}  \text{Pe}^q \, \ket{\psi_{n,m,s}^{\text{R},(q)}} \nonumber \\ 
& = \sum_{q=0}^{\infty} \text{Pe}^q \, \lambda_{n,m,s}^{(q)} \sum_{q'=0}^{\infty} \text{Pe}^{q'} \, \ket{\psi_{n,m,s}^{\text{R},(q')}} \; .
\end{align}
The zero-order term of the previous equation corresponds to the eigenvalue problem of the reference passive system.
Considering the first-order term and operating on the right with $\bra{\psi_{n,m,s}}$ we obtain $\lambda_{n,m,s}^{(1)}=0$.
Inserting the above equations in the expression of the propagator, Eq.~\eqref{eq:solution_propagator_full_problem}, we then have
\begin{align}\label{eq:solution_propagator_full_problem_firstorder}
\mathbb{P} &(\vec{r}, \vartheta, t | \vec{r}_0 , \vartheta_0) = \mathbb{P}^{(0)} (\vec{r}, \vartheta, t | \vec{r}_0 , \vartheta_0)  \nonumber \\
& + \text{Pe} \, p(x) \sum_{n,m,s} e^{-\lambda_{n,m,s}t}  \big[ \psi_{n,m,s}(\vec{r}_0 ,\vartheta_0)^* \, \psi_{n,m,s}^{\text{R},(1)}(\vec{r} ,\vartheta) \nonumber \\ 
    & \qquad + \psi_{n,m,s}^{\text{L},(1)}(\vec{r}_0 ,\vartheta_0)^* \, \psi_{n,m,s}(\vec{r} ,\vartheta) \big] + O(\text{Pe}^2) \; .
\end{align}
We now observe that the eigenvalues spectrum, Eq.~\eqref{eq:EVa}, is degenerate because $\lambda_{n,m,s} = \lambda_{n,-m,s} = \lambda_{n,m,-s} = \lambda_{n,-m,-s}$.
In the following, we assume that these are the only cases of degeneracy.
Following standard perturbation theory~\cite{SakuraiQM} in presence of a degenerate spectrum, we then have
\begin{align}
\ket{\psi_{n,m,s}^{\text{R},(1)}} & =\!\!\! \!\!\! \!\!\!
\sum_{\stackrel{n',m',s'}{\scriptscriptstyle \lambda_{n',m',s'} \neq \lambda_{n,m,s}}} \!\!\!\!\!\! \dfrac{\bra{\psi_{n',m',s'}}\mathcal{L}_1\ket{\psi_{n,m,s}}}{\lambda_{n,m,s} - \lambda_{n',m',s'}} \ket{\psi_{n',m',s'}} \nonumber \\
= -\dfrac{d}{\tau} \Bigg\{  \dfrac{1}{2}  \! \sum_{n'=0}^{\infty} & \! b_{n,n'} \Big( \dfrac{\ket{\psi_{n',m,s+1}}}{\lambda_{n,m,s} \!-\! \lambda_{n',m,s+1}} \!+\! \dfrac{\ket{\psi_{n',m,s-1}}}{\lambda_{n,m,s} \!-\! \lambda_{n',m,s-1}} \Big) \nonumber \\
 - \dfrac{m \pi i}{4 \alpha d} \!  \sum_{m'=0}^{\infty} & \! a_{m,m'} \Big( \dfrac{\ket{\psi_{n,m',s+1}}}{\lambda_{n,m,s} \!-\! \lambda_{n,m',s+1}} \!-\! \dfrac{\ket{\psi_{n,m',s-1}}}{\lambda_{n,m,s} \!-\! \lambda_{n,m',s-1}} \Big) \!\!  \Bigg\} \; .
\end{align}
Similarly, for the left perturbed eigenvectors we have
\begin{align}
\bra{\psi_{n,m,s}^{\text{L},(1)}} & =\!\!\! \!\!\! \!\!\!
\sum_{\stackrel{n',m',s'}{\scriptscriptstyle \lambda_{n',m',s'} \neq \lambda_{n,m,s}}} \!\!\!\!\!\! \dfrac{\bra{\psi_{n,m,s}}\mathcal{L}_1\ket{\psi_{n',m',s'}}}{\lambda_{n,m,s} - \lambda_{n',m',s'}} \bra{\psi_{n',m',s'}} \nonumber \\
= -\dfrac{d}{\tau} \Bigg\{  \dfrac{1}{2} \! \sum_{n'=0}^{\infty} & \! b_{n',n} \Big( \dfrac{\bra{\psi_{n',m,s-1}}}{\lambda_{n,m,s} \!-\! \lambda_{n',m,s-1}} \!+\! \dfrac{\bra{\psi_{n',m,s+1}}}{\lambda_{n,m,s} \!-\! \lambda_{n',m,s+1}} \Big) \nonumber \\
 - \dfrac{m \pi i}{4 \alpha d} \!  \sum_{m'=0}^{\infty} & \! a_{m',m} \Big( \dfrac{\bra{\psi_{n,m',s-1}}}{\lambda_{n,m,s} \!-\! \lambda_{n,m',s-1}} \!-\! \dfrac{\bra{\psi_{n,m',s+1}}}{\lambda_{n,m,s} \!-\! \lambda_{n,m',s+1}} \Big) \!\! \Bigg\} \; .
\end{align}
By exploiting the perturbation theory formalism, higher-order terms in the P\'eclet number can, in principle, be worked out analytically. 
However, such a calculation becomes increasingly tedious.

\section*{Survival probability and first-passage-time distribution}

The knowledge of the propagator allows to compute also the survival probability at time $t$.
The latter, given some initial conditions $(\vec{r}_0, \vartheta_0)$, is readily obtained by integrating over the final position and orientation
\begin{align}\label{eq:survival_probability}
S(t|\vec{r}_0, \vartheta_0) \! = \! \int_{-d}^d \!\!\! \diff x \! \int_0^{2 \alpha d} \!\!\! \diff y \int_0^{2\pi} \!\!\!\! \diff \vartheta \, \mathbb{P}(\vec{r},\vartheta,t|\vec{r}_0,\vartheta_0) \; .
\end{align}
Since 
\begin{align}
\int_{-d}^d \!\!\! \diff x \! \int_0^{2 \alpha d} \!\!\! \diff y \int_0^{2\pi} \!\!\!\! \diff \vartheta  \, p(x) \psi_{n,m,s}(\vec{r},\vartheta) = \dfrac{4\delta_{s,0}}{\pi}   h_m f_n  \; ,
\end{align}
with $h_m := 0$ if $m$ is even, $h_m : 1/m$ if $m$ is odd, and
\begin{align}
f_{n} \! := \! \left\lbrace  \!\!\!
\begin{array}{l}
\dfrac{1}{\mathcal{N}_n} \displaystyle \!\! \int_{-d}^d \!\!\! \diff x \; {_1\!}F_1 \!\! \left(  \dfrac{1}{2} \!-\! \dfrac{\sigma_n}{2} \!; 
\dfrac{1}{2} \!; \dfrac{\beta k x^2}{2} \right)  \mbox{  if } n=0,2,\ldots \\
\\
 0  \qquad \qquad \mbox{else} \, ,
\end{array}
\right. 
\end{align}
exploiting Eqs.~\eqref{eq:solution_propagator_full_problem},~\eqref{eq:decomposition_eigenstates1}, and~\eqref{eq:decomposition_eigenstates2}, one obtains
\begin{align}\label{eq:survival_probability2}
   S&(t|\vec{r}_0, \vartheta_0)  = \dfrac{4}{\pi}  \sum_{n,m,s} e^{-\lambda_{n,m,s}^{\text{Pe}}t} \nonumber \\
   & \!\!\!\!\! \times \!\!\!\!\! \sum_{n',m',s'} \!\!\! g_{n,m,s}^{\text{L};\, n',m',s'} \psi_{n',m',s'}(\vec{r}_0 ,\vartheta_0)^* \!\!\! \sum_{n'',m''} \!\!\! g_{n,m,s}^{\text{R};\, n'',m'',0}  h_{m''}  f_{n''} \, .
\end{align}

\begin{figure}[t!]
\centering
\includegraphics[scale=1]{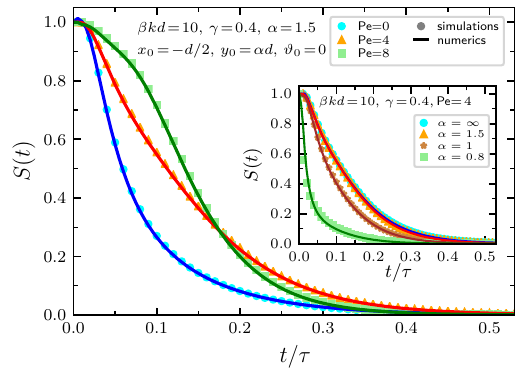}
\caption{Survival probability, $S(t) = S(t| \vec{r}_0,\vartheta_0)$, as a function of time for different $\text{Pe}$ and with initial condition $x_0=-d/2$, $y_0=\alpha d$, and $\vartheta_0 = 0$.
Comparison between simulations (symbols) and numerics (lines) for $\beta k d^2 =10$, $\alpha=1.5$, and $\gamma=0.4$
For the simulations, statistics has been collected from $10^5$ independent particles.
For the numerics, $n_{\rm max}=16$, $m_{\rm max}=14$, and $s_{\rm max}=12$.
Inset: Survival probability for different $\alpha$ at $\text{Pe}=4$, $\beta k d^2 =10$, and $\gamma=0.4$, and with initial condition $x_0=-d/2$, $y_0=\alpha d$, and $\vartheta_0 = 0$. The curve for $\alpha=\infty$ is obtained from the theory developed in Ref.~\cite{Caraglio2025}.
\label{fig:survival_probability}}
\end{figure}

See Fig.~\ref{fig:survival_probability} for a comparison between the results at different P{\'e}clet numbers obtained by numerics and by direct stochastic simulations.
As expected, due to their activity and the persistence of their motion, active particles display a different behavior with respect to standard passive Brownian particles.
For example, starting with an initial position at $x_0=-d/2$ and $y_0=\alpha d$, we can note that a passive particle has a larger probability of being absorbed at shorter times in comparison to an ABP with initial direction pointing towards the barrier ($\vartheta_0=0$) because the latter has higher chances of surviving longer while climbing the potential barrier, see Fig.~\ref{fig:survival_probability}.
However, once the active particle reaches the peak of the energy potential, the self-propulsion starts enhancing the probability of being absorbed because it promotes a quicker reach of the right boundary, see also the inset of Fig.~5 in Ref.~\cite{Caraglio2025}.
As expected, the shorter the size of the domain perpendicular to the barrier, the faster the decay of the survival probability. 
Given $\beta k d^2 =10$, $\gamma=0.4$, and the above specified initial conditions, while with $\alpha=1.5$ the survival is very similar to the one obtained without absorbing boundaries at $y=0,2\alpha d$ and has a halving time of about $0.12\tau$, with $\alpha=0.8$ the particle halves its survival in a time of about $0.07\tau$, see inset of Fig.~\ref{fig:survival_probability}.

Taking the expasion of the propagator at the first order, Eq.~\eqref{eq:solution_propagator_full_problem_firstorder}, the survival reads
\begin{align}\label{eq:survival_probability_first_order}
   S(t|\vec{r}_0, \vartheta_0) \!=\! S^{(0)}(t|\vec{r}_0, \vartheta_0) \!+\! \text{Pe} \, S^{(1)}(t|\vec{r}_0, \vartheta_0) \!+\!  O(\text{Pe}^2) \; ,
\end{align}
with 
\begin{align}\label{eq:survival_probability_first_order_0}
   S^{(0)}(t|\vec{r}_0, \vartheta_0)  = \dfrac{4}{\pi} \sum_{n,m} e^{-\lambda_{n,m,0}t} h_m f_n \psi_{n,m,0}(\vec{r}_0 ,\vartheta_0)^* \; ,
\end{align}
and
\begin{align}\label{eq:survival_probability_first_order_1}
   S^{(1)}&(t|\vec{r}_0, \vartheta_0)  = \dfrac{4}{\pi} \sum_{n,m} \nonumber \Bigg[ -\dfrac{d}{2\tau} e^{-\lambda_{n,m,-1}t} \psi_{n,m,-1}(\vec{r}_0 ,\vartheta_0)^* \\
       \times \Bigg( & h_m \sum_{n'} \! \dfrac{ b_{n,n'} f_{n'}}{\lambda_{n,m,-1} \!-\! \lambda_{n',m,0}}  \!-\! \dfrac{m \pi i}{2 \alpha d } f_n \sum_{m'} \dfrac{ a_{m,m'} h_{m'}}{\lambda_{n,m,-1} \!-\! \lambda_{n,m',0}} \Bigg) \nonumber \\
   & \qquad \qquad - \dfrac{d}{2\tau} e^{-\lambda_{n,m,1}t} \psi_{n,m,1}(\vec{r}_0 ,\vartheta_0)^*   \nonumber \\
    \times \Bigg( & h_m \sum_{n'} \! \dfrac{ b_{n,n'} f_{n'}}{\lambda_{n,m,1} \!-\! \lambda_{n',m,0}} + \dfrac{m \pi i}{2 \alpha d } f_n \sum_{m'} \dfrac{ a_{m,m'} h_{m'}}{\lambda_{n,m,1} \!-\! \lambda_{n,m',0}} \Bigg) \nonumber \\
   &  \qquad \qquad + e^{-\lambda_{n,m,0}t} \psi_{n,m,0}^{L,(1)}(\vec{r}_0 ,\vartheta_0)^*  h_m f_n    \Bigg]  \; .
\end{align}

\begin{figure}[t!]
\centering
\includegraphics[scale=1]{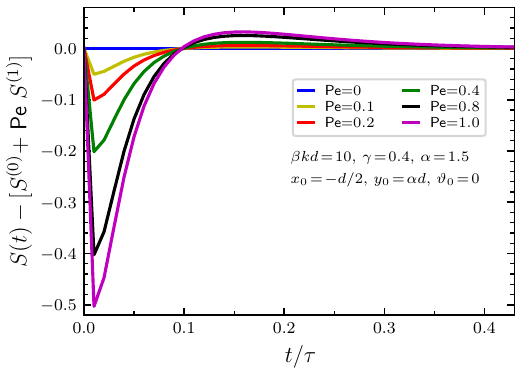}
\caption{For different P\'eclet numbers, difference between the survival probability, $S(t) = S(t| \vec{r}_0,\vartheta_0)$ computed numerically according to Eq.~\eqref{eq:survival_probability2}, and its value at the leading order in $\text{Pe}$, $S^{(0)}(t) + \text{Pe} \, S^{(1)}(t)$, as given in Eqs.~\eqref{eq:survival_probability_first_order_0} and~\eqref{eq:survival_probability_first_order_1}.
Initial condition: $x_0=-d/2$, $y_0=\alpha d$, and $\vartheta_0 = 0$. Other parameters: $\beta k d^2 =10$, $\gamma=0.4$, and $\alpha=1.5$.
For the numerics, $n_{\rm max}=16$, $m_{\rm max}=14$, and $s_{\rm max}=12$.
\label{fig:delta_survival_probability}}
\end{figure}

The difference between the full survival probability computed numerically according to Eq.~\eqref{eq:survival_probability2} and its value at leading order in $\text{Pe}$ is shown in Fig.~\ref{fig:delta_survival_probability}. 
As expected, the first-order expansion of the survival probability becomes less and less accurate with increasing P\'eclet number. 
In particular, for $t \lesssim 0.1 \tau$, it overestimates the true value, even exceeding unity for very small times, while for $t \gtrsim 0.1 \tau$ it underestimates the correct value.

\begin{figure}[t!]
\centering
\includegraphics[scale=1]{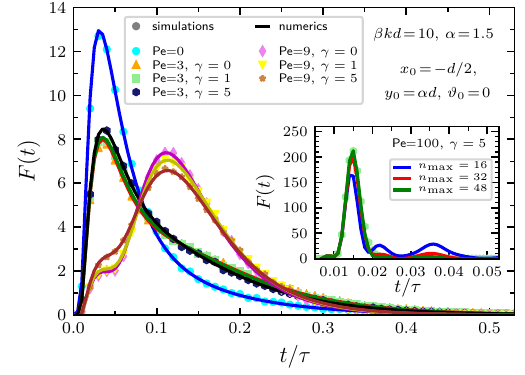}
\caption{First-passage-time distribution, $F(t) = F(t| \vec{r}_0,\vartheta_0)$ for different $\text{Pe}$ and $\gamma$ and with initial condition $x_0=-d/2$, $y_0=\alpha d$, and $\vartheta_0 = 0$.
Comparison between simulations (symbols) and numerics (lines) for $\beta k d^2 =10$ and $\alpha=1.5$.
For the simulations, statistics has been collected from $10^7$ independent particles.
For the numerics, $n_{\rm max}=32$, $m_{\rm max}=30$, and $s_{\rm max}=6$.
In the inset the same quantity is reported for $\text{Pe}=100$ and $\gamma=5$. Here, $m_{\rm max}=12$, $s_{\rm max}=6$, and $n_{\rm max}$ changes according to the caption.
\label{fig:fpt}}
\end{figure}

Finally, starting from Eq.~\eqref{eq:survival_probability} we can also obtain the first-passage-time distribution for any given initial condition as
\begin{align}\label{eq:first_passage_def}
F(t|\vec{r}_0, \vartheta_0) = - \dfrac{\diff S(t|\vec{r}_0, \vartheta_0)}{\diff t} \; .
\end{align}
As for the survival probability, the ABP exhibits first-passage properties that differ from those of a passive particle.
For example, again starting with the initial state ($x_0=-d/2$, $y_0=\alpha d$, $\vartheta_0=0$), when comparing the first-passage-time distribution at large P{\'e}clet numbers ($\text{Pe}=9$) with that of the passive case ($\text{Pe}=0$) or at small P{\'e}clet numbers ($\text{Pe}=3$), the former shows an inflection at $t \approx 0.05\tau$ and a faster decay at longer times, see Fig.~\ref{fig:fpt}.
Thus, similar to what observed in other studies concerning first-passage-time properties of active system~\cite{DiTrapani2023,Malakar2018,Basu2018,Woillez2019} the nonequilibrium character of the dynamics induces peculiarities not displayed by passive particles.
One can also note that, at least for the considered initial conditions and parameters, the rotational diffusivity influences the shape of distribution only to a limited extent, see Fig.~\ref{fig:fpt}.

\begin{figure}[t!]
\centering
\includegraphics[scale=1]{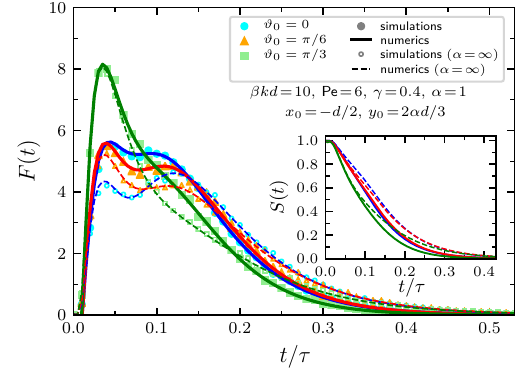}
\caption{
First-passage-time distribution, $F(t) = F(t| \vec{r}_0,\vartheta_0)$ with initial condition $x_0=-d/2$, $y_0=2\alpha d/3$, and different $\vartheta_0$'s.
Comparison between simulations (symbols) and numerics (lines) for $\beta k d^2 =10$, $\text{Pe}=6$, $\gamma=0.4$, and $\alpha=1$.
For the simulations, statistics has been collected from $10^7$ independent particles.
For the numerics, $n_{\rm max}=60$, $m_{\rm max}=12$, and $s_{\rm max}=6$.
Empty symbols (simulations) and dashed lines (numerics) report the corresponding distribution for $\alpha=\infty$.
The curves for $\alpha=\infty$ are obtained from the theory developed in Ref.~\cite{Caraglio2025}.
Inset: Corresponding survival probabily (numerics).
\label{fig:fpt2}}
\end{figure}

Furthermore, the inset of Fig.~\ref{fig:fpt} reports the first-passage-time distribution in the case of an extreme value of the activity, $\text{Pe}=100$.
In this case, we show that more eigenfunctions have to be considered to reach a satisfying solution. 
This is particularly due to the fact that, given the large $\text{Pe}$ value, the first-passage-time distribution is different from zero only at very small times, where the decaying exponential appearing in Eq.~(\ref{eq:solution_propagator_full_problem}) plays a smaller role, meaning that more eigenfunctions have to be considered in the summation providing the propagator.

As expected, the first-passage-time distribution obtained in the present environment differs from that obtained when the absorbing boundaries perpendicular to the barrier are removed. 
For example Fig.~\ref{fig:fpt2} reports the case at P\'eclet number $\text{Pe}=6$ and rotationality $\gamma = 0.4$, and with a form factor $\alpha=1$, starting with the initial position ($x_0=-d/2$, $y_0=2\alpha d/3$) and with different initial self-propulsion directions $\vartheta_0$.
One can observe that for times between approximately $0.02\tau$ and $0.2\tau$, the first-passage probability is higher when the absorbing boundaries at $y = 0$ and $y = 2\alpha d$ are present, while, because of normalization, the opposite holds at longer times, see Fig.~\ref{fig:fpt2}.
Interestingly, this feature is even more pronounced when the self-propulsion direction initially points directly toward the boundary at $x = d$.
Furthermore, in this case, the first-passage-time distribution is bimodal.
These observations can be better understood by examining the survival probability, shown in the inset of Fig.~\ref{fig:fpt2}.
As expected, the survival probability decays more slowly when the absorbing boundaries perpendicular to the barrier are absent.
However, its detailed shape is such that the derivative operation in Eq.~\eqref{eq:first_passage_def} gives rise to the non-trivial features observed in the first-passage-time distribution.

\begin{figure}[t!]
\centering
\includegraphics[scale=1]{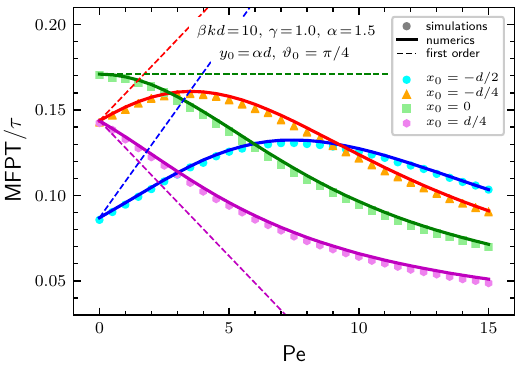}
\caption{Mean first-passage time (MFPT) as a function of the P\'eclet number for different initial $x_0$'s with $y_0=\alpha d$, and $\vartheta_0 = \pi/4$.
Comparison between simulations (symbols) and numerics (lines) for $\beta k d^2 =10$, $\gamma=1.0$, and $\alpha=1.5$.
First-order approximation in $\text{Pe}$ is reporded with dashed lines.
For the simulations, statistics has been collected from $10^5$ independent particles.
For the numerics, $n_{\rm max}=48$, $m_{\rm max}=6$, and $s_{\rm max}=4$.
\label{fig:mfpt}}
\end{figure}

Having the first-passage-time distribution one can also calculate the mean first-passage time (MFPT)
\begin{align}\label{eq:mean_first_passage_def}
\text{MFPT} = \int_0^{\infty} \diff t\, t\, F(t|\vec{r}_0, \vartheta_0) = \int_0^{\infty} \diff t \, S(t|\vec{r}_0, \vartheta_0) \; .
\end{align}
When the $x$ component of the self-propulsion direction points toward the barrier, this observable displays a non-monotonic behavior with increasing P\'eclet number, see Fig.~\ref{fig:mfpt}. 
For example, if the particle is initially placed on the left side of the barrier, a small activity restrains its sliding to the left and consequent absorption at the left boundary, whereas a large activity enhances the chances of quickly crossing the barrier and being absorbed at the right boundary. 
On the other hand, as expected, if the $x$ component of the self-propulsion points opposite to the gradient of the potential, the MFPT monotonically decreases with increasing P\'eclet number.
The latter considerations can be made more quantitative by examining the probability of absorption at the right ($x=d$) and left ($x=-d$) boundaries, as discussed in the next section.
Finally, we note that the first-order approximation of the MFPT in the P\'eclet number is linear, as expected when integrating $S^{(0)}(t) + \text{Pe} , S^{(1)}(t)$ over time, and it becomes rapidly inaccurate with increasing activity.

\section*{Probability of absorption at a given boundary}

The continuity equation $\partial_t \mathbb{P} = - \vec{\nabla} \cdot \vec{j}$~\cite{note2} allows us to write particle current in the $x$ and $y$ directions associated to the Fokker-Planck equation
\begin{align}\label{eq:current_x}
j_x ( \vec{r},\! \vartheta, t  | \vec{r}_0, \! \vartheta_0)  \!=\!  \dfrac{d^2}{\tau} \!\! \left[  \beta k x \!-\! \partial_x \!+\!  \dfrac{\text{Pe}}{d}  \cos \vartheta \right]   \mathbb{P} (\vec{r}, \! \vartheta, t | \vec{r}_0, \! \vartheta_0) \, ,
\end{align}
and
\begin{align}\label{eq:current_y}
j_y ( \vec{r},\! \vartheta, t  | \vec{r}_0, \! \vartheta_0)  \!=\!  \dfrac{d^2}{\tau} \!\! \left[  - \partial_y \!+\!  \dfrac{\text{Pe}}{d}  \sin \vartheta \right]   \mathbb{P} (\vec{r}, \! \vartheta, t | \vec{r}_0, \! \vartheta_0) \, .
\end{align}
Starting from these expressions we can calculate the probability of absorption at a given boundary conditioned to the initial conditions $(\vec{r}_0,\vartheta_0)$.
In particular, we have that the probability of being absorbed on the boundary at $x=d$ is
\begin{align}\label{eq:right_absorbtion}
P^{\rm absorb}_{x=d} (\vec{r}_0, \! \vartheta_0) \!=\! \int_0^{\infty} \!\! \diff t \int_0^{2 \alpha d} \!\!\! \diff y \int_0^{2\pi} \!\!\! \diff \vartheta \, j_x ( d,y,\! \vartheta, t  | \vec{r}_0, \! \vartheta_0) \, .
\end{align}
Considering that $\psi_{n,m,s}(x=d,y,\vartheta)=0$ and inserting Eqs.~\eqref{eq:solution_propagator_full_problem},~\eqref{eq:peq},~\eqref{eq:decomposition_eigenstates1}, and~\eqref{eq:decomposition_eigenstates2} into Eq.~\eqref{eq:right_absorbtion} we get
\begin{align}\label{eq:right_absorbtion_2}
P^{\rm absorb}_{x=d} & (\vec{r}_0, \! \vartheta_0)  \!=\! -\dfrac{4 d^2}{\pi \tau} e^{\beta k d^2/2} \sum_{n,m,s} \dfrac{1}{\lambda_{n,m,s}^{\text{Pe}}} \, \psi_{n,m,s}^{\text{Pe},\text{L}}(\vec{r}_0 ,\vartheta_0)^* \nonumber \\
& \times \sum_{n',m'} g_{n,m,s}^{\text{R};\, n',m',0} \, h_{m'} \, \partial_x \mathcal{X}_{n'}(x=d) \; .
\end{align}

Similarly we have
\begin{align}\label{eq:left_absorbtion}
P^{\rm absorb}_{x=-d} & (\vec{r}_0, \! \vartheta_0) \!=\! - \! \int_0^{\infty} \!\!\! \diff t \int_0^{2 \alpha d} \!\!\!\! \diff y \int_0^{2\pi} \!\!\!\! \diff \vartheta \, j_x ( -d,y,\! \vartheta, t  | \vec{r}_0, \! \vartheta_0) \nonumber \\
 & = \dfrac{4 d^2}{\pi \tau} e^{\beta k d^2/2} \sum_{n,m,s} \dfrac{1}{\lambda_{n,m,s}^{\text{Pe}}} \, \psi_{n,m,s}^{\text{Pe},\text{L}}(\vec{r}_0 ,\vartheta_0)^* \nonumber \\
& \quad  \times \sum_{n',m'} g_{n,m,s}^{\text{R};\, n',m',0} \, h_{m'} \, \partial_x \mathcal{X}_{n'}(x=-d) \; ,
\end{align}
\begin{align}\label{eq:upper_absorbtion}
P^{\rm absorb}_{y=2 \alpha d} & (\vec{r}_0, \! \vartheta_0) \!=\!  \int_0^{\infty} \!\!\! \diff t \int_{-d}^{d} \!\! \diff x \int_0^{2\pi} \!\!\!\! \diff \vartheta \, j_y ( x,2\alpha d,\! \vartheta, t  | \vec{r}_0, \! \vartheta_0) \nonumber \\
 & = - \dfrac{\pi}{2 \alpha^2 \tau} \sum_{n,m,s} \dfrac{1}{\lambda_{n,m,s}^{\text{Pe}}} \, \psi_{n,m,s}^{\text{Pe},\text{L}}(\vec{r}_0 ,\vartheta_0)^* \nonumber \\
& \quad  \times \sum_{n',m'} g_{n,m,s}^{\text{R};\, n',m',0} \, f_{n'} \, m' \cos(m' \pi) \; ,
\end{align}
and
\begin{align}\label{eq:lower_absorbtion}
P^{\rm absorb}_{y=0} & (\vec{r}_0, \! \vartheta_0) \!=\!  - \! \int_0^{\infty} \!\!\! \diff t \int_{-d}^{d} \!\! \diff x \int_0^{2\pi} \!\!\!\! \diff \vartheta \, j_y ( x,0,\! \vartheta, t  | \vec{r}_0, \! \vartheta_0) \nonumber \\
 & =  \dfrac{\pi}{2 \alpha^2 \tau} \sum_{n,m,s} \dfrac{1}{\lambda_{n,m,s}^{\text{Pe}}} \, \psi_{n,m,s}^{\text{Pe},\text{L}}(\vec{r}_0 ,\vartheta_0)^* \nonumber \\
& \quad  \times \sum_{n',m'} g_{n,m,s}^{\text{R};\, n',m',0} \, f_{n'} \, m' \; .
\end{align}

\begin{figure}[t!]
\centering
\includegraphics[scale=1]{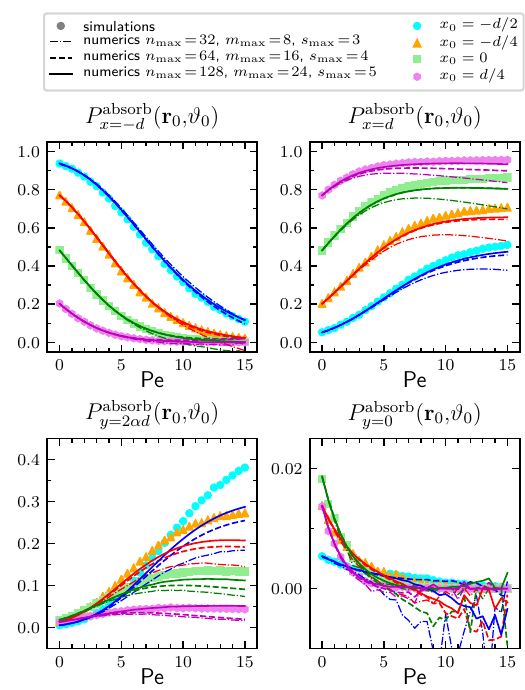}
\caption{Absorption probability at different boundaries as a function of the P\'eclet number for different initial $x_0$'s and with $y_0=\alpha d$, and $\vartheta_0 = \pi/4$.
Comparison between simulations (symbols) and numerics (lines) for $\beta k d^2 =10$, $\gamma=1.0$, and $\alpha=1.5$.
For the simulations, statistics has been collected from $10^5$ independent particles.
For the numerics, full line corresponds to $n_{\rm max}=128$, $m_{\rm max}=24$, and $s_{\rm max}=5$; dashed line to $n_{\rm max}=64$, $m_{\rm max}=16$, and $s_{\rm max}=4$; dot-dashed line to $n_{\rm max}=32$, $m_{\rm max}=8$, and $s_{\rm max}=3$.
\label{fig:absp}}
\end{figure}

The absorption probabilities at the different boundaries as a function of the P\'eclet number, for various initial conditions, are reported in Fig.~\ref{fig:absp}.
As expected, for the considered initial conditions (all characterized by $y_0 = \alpha d$ and $\vartheta_0 = \pi/4$) the probability of being absorbed at the boundaries located at $x = d$ or $y = 2\alpha d$ increases with increasing activity, while the probability of absorption at the remaining boundaries correspondingly decreases.
Interestingly, we note that the numerical evaluation of the absorption probabilities becomes increasingly challenging as the P\'eclet number increases, particularly in the case of absorption on the upper boundary (at $y=2\alpha d$).
This difficulty stems from the fact that in Eqs.~(\eqref{eq:right_absorbtion}--\eqref{eq:lower_absorbtion}), the perturbed eigenvalues appear in the form $1/\lambda$, rather than through factors of the type $\exp(-\lambda t)$ as in the expressions of the propagator, Eq.~\eqref{eq:solution_propagator_full_problem}, or of the survival probability, Eq.~\eqref{eq:survival_probability2}.
As a consequence, a larger number of terms in the series expansions of Eqs.~(\eqref{eq:right_absorbtion}--\eqref{eq:lower_absorbtion}) provide non-negligible contributions to the final result.
Unfortunately, we could not test our numerics for values of $n_{\rm max}$, $m_{\rm max}$, and $s_{\rm max}$ larger than those reported in Fig.~\ref{fig:absp}.
This limitation arises from the need to diagonalize the full operator $\mathcal{L}$, which requires $O(N^{3})$ computational operations and $O(N^{2})$ memory storage, where $N=(1+n_{\rm max})m_{\rm max} (2s_{\rm max} +1)$ is the dimension of the truncated operator.

\section*{Conclusions}

We have derived an exact series solution for the probability propagator of a two-dimensional ABP living in a rectangular domain with an absorbing boundary and subjected to a parabolic potential along the $x$ direction.
Taking inspiration from recent literature~\cite{Caraglio2022,DiTrapani2023}, the method to obtain such a solution requires the knowledge of the eigenvalues and eigenfunctions of a reference system given by the standard passive Brownian particle and then deal with the activity of the particle as a perturbation of this reference system.
The propagator is then expressed in terms of the perturbed left and right eigenvectors, which can be easily computed by direct diagonalization of the matrix form of
the Fokker-Planck operator, multiplied by an exponentially decaying factor with a rate given by the corresponding perturbed eigenvalue.
Note that, while perturbation theories generally lead to approximate solutions~\cite{SakuraiQM}, here, once the perturbed eigenfunction and eigenvalues are found, one can directly get the propagator of the ABP as in Eq.~(\ref{eq:solution_propagator_full_problem}). 
In this sense, in our theory the solution is exact, with the only limitation that to make the numerics possible, we have to cut off the number of considered eigenfunctions.
However, as done in previous work~\cite{DiTrapani2023}, also in the current case one can fully exploit the formalism of perturbation theory and express the solution as a power series in the P\'eclet number, $\text{Pe}$, but obviously, truncating this series would result in an approximated solution.
For comparison, throughout the manuscript we have also evaluated several observables at the first leading order in the P\'eclet number.

By integrating over the proper variables, the knowledge of the propagator is then exploited to compute the spatial probability density at a later time, the survival probability, the first-passage-time distribution, the mean first-passage time, and the absorption propbability at a given boundary.
In line with what observed in the case of an infinite domain in the direction perpendicular to the barrier~\cite{Caraglio2025} and of a circular absorbing boundary~\cite{DiTrapani2023}, these observables show a strong dependence on the activity of the particle and, to a lesser extent, on its rotational diffusivity.
We also note that, once the propagator is known, one can also compute moments and correlation functions~\cite{Caraglio2022}. 
However, in the present case involving absorbing boundaries, such observables are well defined only as long as the survival probability remains unity.

Our findings extend the nowadays limited set of exactly solvable models for active particles~\cite{Wagner2017,Hermann2018,Schnitzer1993,Tailleur2008, Tailleur2009,Malakar2018,Kurzthaler2016,Kurzthaler2017,Kurzthaler2018,Martens2012,Sevilla2015,Caraglio2022,DiTrapani2023,Caraglio2025} and the knowledge of first-passage properties of active particles~\cite{Malakar2018,Angelani2014,Scacchi2018,Demaerel2018,Dhar2019,Basu2018,Caprini2021}.
They can also be exploited to reach analytical insight into target-search problems in complex environments involving absorbing boundaries and potential barriers~\cite{Tejedor2012,Volpe2017,Zanovello2021,Zanovello2021b}.

Finally, it is worth mentioning that, in principle, it is easy to solve the problem of a rectangular domain with an absorbing boundary also in the absence of any energy barrier.
In fact, in this case,  both $x$- and $y$-dependent parts of the eigenfunctions of the reference passive system have the same sinusoidal behavior as that of the $y$-dependent part in the current problem and, starting from this observation, one can readily obtain the eigenvalues of the reference passive system and the action of the full operator on such eigenfunctions.
Similarly, in principle, one can also consider the case $k<0$, which corresponds to a harmonic potential along the $x$-axis. 
Although, in doing so, one must properly compute the eigenvalues, taking into account that the Kummer confluent hypergeometric function satisfies the property $\,_1F_1(a;b;z) = e^z \,_1F_1(b-a;b;-z)$.
In the case of a harmonic potential along the $x$-axis, the limit of an infinite distance between the left and right absorbing boundaries is particularly interesting. 
However, in this limit, it is convenient to adopt a different definition of the length unit, namely the thermal oscillation length $\sqrt{k_BT/|k|}$, and the $x$-dependent part of the eigenfunctions becomes a Bessel function.
A comprehensive discussion of this case will be reported in a separate work currently in preparation.

\begin{acknowledgments}
M.C. is supported by FWF: P 35872-N and acknowledges Thomas Franosch and Enrico Carlon for fruitful discussions.
\end{acknowledgments}
\medskip

\appendix

\begin{widetext}

\section{Action of $\mathcal{L}_1$ on the passive system's eigenfunction $\ket{\psi_{n,m,s}}$} \label{sec_appA}
Here, we now explicitly evaluate the action of the perturbation $\mathcal{L}_1$ on the eigenstates of $\mathcal{L}_0$ obtained from the eigenvalue problem
\begin{align}
    \mathcal{L}_0 \ket{\psi_{n,m,s}} = -\lambda_{n,m,s} \ket{\psi_{n,m,s}} \; .
\end{align}
Considering Eqs.~\eqref{eq:L1} and~\eqref{eq:eigenfunctions} we have
\begin{align} \label{eq:L1action_App}
\mathcal{L}_1  & \ket{\psi_{n,m,s}}  = -\dfrac{d}{\tau} \Bigg\{  \left( \dfrac{e^{i(s+1)\vartheta} + e^{i(s-1)\vartheta}}{2} \right)  \sin \left(  m \pi \dfrac{ y}{2 \alpha d} \right) (\beta k x + \partial_x) \dfrac{e^{-\beta k x^2/4} Y_n(x)}{\mathcal{N}_n} \nonumber \\
   & \qquad \qquad \qquad \quad - \dfrac{m \pi i}{2 \alpha d} \left( \dfrac{e^{i(s+1)\vartheta} - e^{i(s-1)\vartheta}}{2} \right)  \cos \left(  m \pi \dfrac{ y}{2 \alpha d} \right) \dfrac{e^{-\beta k x^2/4} Y_n(x)}{\mathcal{N}_n}    \Bigg\} \; .
\end{align}

We start by noting that given a complete orthogonal system of functions $\{ \phi_{\ell}(z) \}$ over the interval $\mathcal{R}$, the functions  $\phi_{\ell}(z)$ satisfy an orthogonality relationship of the form 
\begin{align} \label{eq:general_orthogonality}
\int_{\mathcal{R}} \diff z\, w(z) \phi_{\ell'}(z) \phi_\ell(z)   = c_\ell \delta_{\ell,\ell'} \; ,
\end{align}
where $w(z)$ is a a weighting function, $c_\ell$ are given constants and $\delta_{\ell,\ell'}$ is the Kronecker delta. 
An arbitrary function $f(z)$ can be written as a series
\begin{align} \label{eq:general_series}
f(z)= \sum_{\ell=0}^{\infty} a_\ell \phi_\ell(z) \; , 
\end{align}
with
\begin{align} \label{eq:general_coefficients}
a_\ell = \dfrac{1}{c_\ell} \int_{\mathcal{R}} \diff z \, w(z)  \phi_{\ell}(z) f(z)  \; . 
\end{align}

Substituting $z$ with $y$ and applying the above relation to the case in which $\phi_m(y) = \sin(m \pi y/2 \alpha d)$, $f(y) = \cos(m' \pi y/2 \alpha d)$, $w(y) = 1/\alpha d$, and $c_{m}=1$ we have
\begin{align}
\cos \left(  m \pi \dfrac{ y}{2 \alpha d} \right) = \sum_{m'=0}^{\infty} a_{m,m'} \sin \left(  m' \pi \dfrac{ y}{2 \alpha d} \right) \; ,
\end{align}
with
\begin{align}
a_{m,m'} = \left\lbrace \!
\begin{array}{ll}
\\
0 & \mbox{  if } m'=m \\
\\
\dfrac{2[   m' \! \cos(m' \pi) \cos(m \pi) \!+\! m  \sin(m' \pi) \sin(m \pi) \!-\! m'  ] }{\pi(m-m')(m+m')} &   \mbox{  otherwise.}
\end{array} 
\right. 
\end{align}
Note that the above series expansion is valid for $y \in (0,2\alpha d)$ but not at the boundaries $(y=0,2 \alpha d)$ where $\cos(m \pi y / 2\alpha d)$ is either equal to $1$ or $-1$, while $\sin(m' \pi y / 2\alpha d)=0$ for each $m'$.

Finally, inserting into Eq.~\eqref{eq:L1action_App} the above relations and those regarding the $x$ component reported in Appendix~A of Ref.~\cite{Caraglio2025}, one gets
\begin{align} \label{eq:L1action_App_2}
\mathcal{L}_1  & \ket{\psi_{n,m,s}}  = -\dfrac{d}{\tau} \Bigg\{  \dfrac{1}{2} \sum_{n'=0}^{\infty} b_{n,n'} \big( \ket{\psi_{n',m,s+1}} + \ket{\psi_{n',m,s-1}}\big) 
- \dfrac{m \pi i}{4 \alpha d} \sum_{m'=0}^{\infty} a_{m,m'} \big( \ket{\psi_{n,m',s+1}} - \ket{\psi_{n,m',s-1}}\big)  \Bigg\} \; ,
\end{align}
with
\begin{align}
 b_{n,n'} = \left\lbrace 
 \begin{array}{ll}
\dfrac{\beta k(1\!-\!\sigma_n)}{\mathcal{N}_n \mathcal{N}_{n'}}  \displaystyle \int_{-d}^d \diff x \, e^{-\beta k x^2/4} Y_{n'}(x) \, x \, {_1\!}F_1 \! \left(  \dfrac{3}{2} \!-\! \dfrac{\sigma_n}{2} ; \dfrac{3}{2} ; \dfrac{\beta k x^2}{2} \right)
    & \mbox{ if } n=0,2,\ldots \mbox{ and } n'=1,3,\ldots \\ 
    & \\
\dfrac{\sqrt{\beta k}}{\mathcal{N}_n \mathcal{N}_{n'}} \displaystyle \int_{-d}^d \diff x \, e^{-\beta k x^2/4} Y_{n'}(x)   \Bigg[ {_1\!}F_1 \! \left( \!  1 \!-\! \dfrac{\sigma_n}{2}  ;  \dfrac{3}{2} ; \dfrac{\beta k x^2}{2} \! \right) & \\
\quad + \dfrac{2-\sigma_n}{3} \beta k x^2 {_1\!}F_1 \! \left(  2 \!-\! \dfrac{\sigma_n}{2} ; \dfrac{5}{2} ; \dfrac{\beta k x^2}{2} \right) \Bigg]
    & \mbox{ if } n=1,3,\ldots \mbox{ and } n'=0,2,\ldots \\
    & \\
 0  & \mbox{ otherwise.} \\
 \end{array}
 \right. 
\end{align}

\section{Details on numerical simulations \label{sec_appB}}

To corroborate our analytical findings, we performed numerical simulations by directly integrating the Langevin equation of motion for a ABP diffusing in a conservative energy landscape, $U(x) = -\tfrac{1}{2}kx^2$. 
When discretized according to the It\^o rule, these equations read
\begin{eqnarray}\label{eom}
\vec{r}_{t\!+\!\Delta t} &=& \vec{r}_{t} + v\, \vec{u}_{t} \, \Delta t - \mu \vec{\nabla} U(\vec{r}_{t}) \Delta t + \sqrt{2D\Delta t} \, \boldsymbol{\xi}_t,\\ \label{eom2}
\vartheta_{t\!+\!\Delta t} &=& \vartheta_{t} + \sqrt{2D_{\rm rot}\Delta t} \, \eta_t,
\end{eqnarray}
where  $\Delta t$ is the integration step, $\vec{r}_t = (x_t,y_t)$ is the position at time $t$ and $\vec{u}_{t} = \big(\cos\vartheta_{t},\sin\vartheta_{t}\big)$ denotes the instantaneous orientation of the driving velocity. 
The components of the vector noise $\boldsymbol{\xi}_t=(\xi_{x,t},\xi_{y,t})$ and of the scalar noise $\eta_t$ are independent random variables, distributed according to a Gaussian with zero average and unit variance.

To compute the various observables reported in the main text, several independent particles starting from the same initial condition are considered. 
When a particle crosses one of the boundaries for the first time, it is considered absorbed.
Note that due to the presence of the absorbing boundary, the discretization time step adopted in the stochastic simulations should be smaller than what is usually adopted for standard simulations of an ABP in free space.
As a matter of fact, with increasing time, convergence of results in the proximity of the boundary becomes more and more sensitive to the value of the discretization time step, see also appendix B in Ref.~\citep{DiTrapani2023}.
Throughout our work, we adopted $\Delta t = 10^{-4}\tau$.

\end{widetext}


\begin{thebibliography}{67}
\bibitem [{not()}]{note2}%
  \BibitemOpen
  \href@noop {} {}\bibinfo {note} {Note that here we abused notation by writing
  $\vec{\nabla}=(\partial_x,\partial_y,\partial_\vartheta)$ and $\vec{j}=(j_x,
  j_y, j_\vartheta)$, thus including in the vectorial notation a $\vartheta$
  component.}\BibitemShut {Stop}%
\end{thebibliography}
\end{document}